# DNA LOSSLESS DIFFERENTIAL COMPRESSION ALGORITHM BASED ON SIMILARITY OF GENOMIC SEQUENCE DATABASE

Heba Afify[1], Muhammad Islam[1] and Manal Abdel Wahed[1]

[1]Department of Systems and Biomedical Engineering, Cairo University, Egypt
hebaaffify@yahoo.com, Manalaw2003@yahoo.com

## ABSTRACT

*Modern biological science produces vast amounts of genomic sequence data. This is fuelling the need for efficient algorithms for sequence compression and analysis. Data compression and the associated techniques coming from information theory are often perceived as being of interest for data communication and storage. In recent years, a substantial effort has been made for the application of textual data compression techniques to various computational biology tasks, ranging from storage and indexing of large datasets to comparison of genomic databases. This paper presents a differential compression algorithm that is based on production of difference sequences according to op-code table in order to optimize the compression of homologous sequences in dataset. Therefore, the stored data are composed of reference sequence, the set of differences, and differences locations, instead of storing each sequence individually. This algorithm does not require a priori knowledge about the statistics of the sequence set. The algorithm was applied to three different datasets of genomic sequences, it achieved up to 195-fold compression rate corresponding to 99.4% space saving.*

## 1. INTRODUCTION

The compression of genomic sequences remains a challenging problem, with profound implications in biology and with important technological impact when the use of genomic data will become a daily practice in health and medicine. As such, it will certainly be investigated further due to several reasons: benefits when storing or transmitting the genome files; possibilities for comparison of entire genomes by similarity metrics approximating Kolmogorov similarity [1, 2]; and discovering statistically significant relationships among various sequences.

Data compression and the related information-theoretic techniques find a wide use for investigation in computational biology. Such a pervasive use has grounds in some outstanding notions that deeply characterizes data compression, in particular universality and quantification of statistical dependence via information measures. These notions give rise to methods that need very few assumptions on the data models and, as a consequence, very minor parameter estimations for their application. This seems to be a major advantage for computational biology applications, where the statistical modeling of the data is a highly non-trivial task. In addition, the low-computational demand of these methods allows them to scale well with dataset size, even on a genomic scale. Obviously, data structure, data modeling and speed are the main advantages for the use of data compression in biological investigations [3].

The amount of DNA being extracted from organisms and sequenced is increasing exponentially [4]. This yields two problems: storage and comprehension. Despite the prevalence of broadband network connections, there still exists a need for compact representation of data to speed up transmission. Transferring a single sequence that is millions of characters long may take ten to fifteen minutes over a dial-up connection. Compression of genomic sequences can decrease the storage requirements, and increase the transmission speed.

Understanding genomic sequences has wide applications, from the synthesis of medicines to genetic screening and engineering. The knowledge of the structure of a sequence is important for its comprehension. If a set of sequences have a common structural property that is shared by another sequence, it is possible that they are related in some way, or that the knowledge that applies to one may also be useful for the other. Compression can help to show both the structure of a sequence and how it is related to other sequences.

DNA is composed of four bases (A, T, G, C), and can be coded using two bits per base. According to functionality, DNA manifests different properties from other kinds of data. Standard compression algorithms for text or image files exploit small repeated patterns and contextual similarities to achieve compression. However, repeated patterns in DNA sequences are typically much longer and less frequent, so standard compression algorithms perform poorly on DNA. The most popular general-purpose encoders of today, such as *gzip* [5], which is based on the Lempel-Ziv algorithm [6], and *bzip2,* based on the Burrows-Wheeler Transform [7], which usually produces more than two bits per base to achieve the un-encoded representation [8]. Hence, the quest for efficient DNA compression programs started to become popular in the competition-driven community of data compression enthusiasts. DNA sequences are compressible, so they are not random. But they are not highly compressible. It is therefore necessary for coding methods to be as efficient as possible. In the context of compression, missing structure will lead to inefficient compression.

In the last two decades, compression of genomic sequences can be divided into two categories: Specific compression algorithms developed for efficiently compressing sequence data for the sake of reduced resource consumption (disk space or network usage) [8-13]; and investigations of the usefulness of compressibility as a measure of information content, for the purpose of making inferences about sequences (such as the relatedness of two sequences) [1, 14]. Specific compression algorithms have been proposed for DNA compression by using particular characteristics such as exact or approximate repeats measures within a single DNA sequence that are based on relations between subsequences only. Compression gains afforded by these algorithms are ultimately not sufficient to justify their adoption for large databases. While much research has been done on compressing individual DNA sequences, surprisingly little has focused on the compression of entire databases [3]. The results of compressing genomic sequences can be applied to the problem of evolution derivation [15]. Compression–based distance measures (CBMs) that depend on probabilities of mismatching locations [16], are not distinct enough among different classes. However, researches have been suffering from the poor modeling to characterize the relationship between sequences.

Algorithms for Compressing DNA sequences, such as GenCompress [11] ,Biocompress [10] and Cfact [17] were available to compress DNA sequences. Their compression rate was about 1.74 bits per base i.e., 78% in compression rate. Hence, a compression algorithm named "GenBit Compress Tool" [18] is presented, whose compression rate was below 1.2 bits per byte (for Best case) , 1.727 bits/bytes (for Average Case), 2.238 bits/bytes (Worst case ) even for larger genome (nearly 2,00,000 characters). Recently, P. Raja Rajeswari, and A. Apparao [19] presented a new compression algorithm named "DNABIT Compress" whose compression rate was below 1.56 bits per base (for Best case) even for larger genome. DNABIT Compress algorithm was the best among the remaining compression algorithms and significantly improves the running time of all previous DNA compression programs.

Researchers have worked in entropy estimation for genomic sequences, either by computing frequency of $n$-mers for long enough inputs, called Shannon entropy [20], or by adopting compression methods to obtain an upper bound on entropy [8]. Loewenstern [21, 22] introduced a compression method CDNA by considering inexact match in finding patterns. Importantly, Lanctot and Yang [23] improved the compression further by exploiting the reverse complement property of DNA sequences. Also, this latter compression method produces a good estimation of entropy, e.g., the estimate approaches the actual entropy for long enough input. Badger and Chen [1] proposed a distance function with nice properties for cluster related sequences.

Interestingly, difference compression schemes targets the efficient compression of entire databases of sequences. DNAzip [24] was first algorithm introducing the important idea of only storing differences to a reference sequence, but in this case for storing an entire, assembled genome as a series of difference. This algorithm [24] did not consider the process of generating the variations, which can be a challenging problem in itself, but assume that the variation data have been provided. Difference compression schemes, which compress entire sets of homologous sequences by encoding only the differences between a genomic sequences and a reference sequence, are also suggested in [25]. Brandon et al [25] found that selecting the reference sequence is important for having an effective compression of dataset. With only a partial level of optimization, 3615 genome sequences occupying 56MB in GenBank are compressed down to only 167 KB, achieving a 345-fold compression rate, using the revised Cambridge Reference Sequence as the reference sequence. Using the consensus sequence as the reference sequence, the data can be stored using only 133 KB, corresponding to a 433-fold level of compression, roughly a 23% improvement. On the other hand, C. Wang [26] implemented a generic tool, GRS, for de novo compression of genome resequencing data which does not need the reference SNPs map. When its performance was tested on the first Korean personal genome sequence dataset, GRS was able to achieve 159-fold compression, reducing the size of the data from 2986.8 to 18.8 MB. While being tested against the sequencing data from rice and Arabidopsis thaliana, GRS compressed the 361.0 MB rice genome data to 4.4 MB, and the A. thaliana genome data from 115.1MB to 6.5 KB. DNAEncodeWG [27] also presented how to compress DNA sequence data using the whole genome sequence of an organism to identify differences between DNA sequences if a repository of the whole genome sequence of the organism is accessible through the Web. It encoded the sequences 10-fold better than the other standard algorithms. Kozanitis [28] focused on fragment compression as opposed to sequence compression by using SLIMGENE. H. Afify [29] presented another algorithm in which for each pair of similar sequences, a third sequence can be generated; representing the difference between them, and the entropy of the generated difference sequence can be estimated. Difference sequence can help in building phylogenic tree, while the entropy can help in selecting appropriate compression reference for short dataset.

For difference compression scheme to be appropriate, it should be suitable for the way the set is being used. When an entire set is archived or transmitted, the concern will be focused on the compression rate. In this situation, a reference sequence is stored, and the next sequence is generated from this reference and the appropriate difference sequence. Newly generated sequences may be used as references for subsequent sequences. When the set is being actively used, in addition to the compression rate, the speed of decompression is of comparable importance, where any sequence may be repeatedly fetched at any moment. To speed up sequence decompression, it is preferable to have a single reference for the whole set.

In this study, we describe another solution to the compression of genomic sequence dataset which compresses the dataset based on comparing it with a reference sequence. We focus our study on large sets of sequences that belong to the same class. If two genomes are, e.g., more than 99% identical, it is much more efficient to store one genome as a variation from the other; in which case, only that 1% representing the variation needs to be stored. A differentially compressed set is a set where a single reference sequence is stored, along with information about the difference between this sequence and the rest of the set. To evaluate the suggested selection method, the compression of the differences, compression of difference locations and the size of the compressed set are examined, as is explained in the next section.

## 2. MATERIAL AND METHOD

### 2.1. Data Extraction

The data used in our work consists of three different datasets of genomic sequences including 3615 human mitochondrial genomic sequences, 500 human virus sequences H1N1, and 100 mouse sequences Mus Musculus Dmesticus.

Mitochondrial dataset takes 56MB size in GenBank, and is downloadable from the GenBank database, HapMap web site and the MITOMAO database [30]. Among the sequences, 2671 correspond to complete genome, while the remaining 944 correspond only to the coding region sequence, which is about 1100bp shorter than the full genome sequence. Virus dataset takes 601KB size in GenBank, and is downloaded from Influenza Research Database [31]. Mouse datasets takes 106KB in GenBank, and is downloaded from Mouse Genome Database (MGD) [32]. References were selected as follows: Cambridge sequence NC_012920 sequence for human mitochondrial dataset, HM17663 for virus dataset, and AJ843867 for mouse dataset.

### 2.2. Operation Code Generation

The DNA is constructed of a double helix held together by hydrogen bonds. Each strand of the helix is a biomolecule consisting of many linked components called nucleotides. Each nucleotide is one of four possible types: adenine (A), cytosine (C), guanine (G) and thymine (T). The two strands of the helix are exact complement of each other. Each nucleotide of one strand matches to its complement on the other strand, where A pairs with T and G pairs with C, DNA strands that are complementary to themselves are called self-complementary or palindromes.

To generate an operation code of differences between a target sequence and its reference, the base at every location of the target is compared with the corresponding bases at the reference. A difference between the two bases can be due to one of three modifications to the target: a base insertion, a base deletion, or a replacement. Table 1 summarizes the operation codes.

Table 1. Operation code generation.

| Corresponding Bases | Operation | Op-codes |
|---|---|---|
| The same | Similarity | "0" |
| A ←→T<br>G ←→C | Replacement | "1" |
| A ←→G<br>C ←→T | | "2" |
| A ←→C<br>G ←→T | | "3" |
| A or T or G or C → '-' | Deletion | "4" |
| '-' →G | Insertion | "5" |
| '-' → A | | "6" |
| '-' → C | | "7" |
| '-' → T | | "8" |

## 2.3. Differential Compression Algorithm

The proposed differential compression algorithm consists of three main steps: Alignment, Differences recording and differences compression.

### 2.3.1. Alignment

We start by aligning each sequence in the dataset with the reference sequence using local Sequence Alignment. The sole purpose of sequence alignments is to place homologous positions of homologous sequences into the same column by inserting gapes. Gaps reflect the occurrence of insertions/deletions or other rearrangements during the process. Also, the alignment of similar sequences can help in discovering patterns and relationships between sequences, consequently improve their compression ratio [33].

### 2.3.2. Differences Recording

In this step, differences and differences locations vectors are recorded. Difference vector is achieved by storing only the op-code of the different bases between the aligned reference sequence and the aligned genomic sequence; these will be according to Table 2 ranging from 1 to 8. Difference locations with respect to the unaligned reference sequence are recorded. It is to be noted out that an insertion in the aligned reference sequence will not change the recorded unaligned reference sequence base location, while any other operation in the aligned genomic sequence will increment to the next recorded unaligned reference sequence base location.

Table 2. Op-codes of differences.

| | | Aligned Reference Sequence Bases | | | | |
|---|---|---|---|---|---|---|
| | | G | A | C | T | - |
| Aligned Genomic Sequence Bases | G | 0 | 2 | 1 | 3 | 5 |
| | A | 2 | 0 | 3 | 1 | 6 |
| | C | 1 | 3 | 0 | 2 | 7 |
| | T | 3 | 1 | 2 | 0 | 8 |
| | - | 4 | 4 | 4 | 4 | 0 |

### 2.3.3. Difference Compression

The proposed compression algorithm can be divided into two phases, the differences are coded, and the codes are compressed in the first phase, while their locations are compressed in the second. To improve the compression of differences locations, the distance between successive locations are stored.

To compare the results obtained after the compression of both differences codes and locations, we used compression ratio which is the ratio between the compressed size and the uncompressed size, and space saving which is defined as the reduction in size relative to the uncompressed size (1-compression ratio) [34], as measures for the compression process. Figure 1 shows the architecture of the proposed differential compression algorithm.

## 3. RESULTS AND DISCUSSION

We have tested the proposed differential compression algorithm on the three genomic sequences datasets. Difference codes were compressed using two lossless compression algorithms namely, Huffman [35] and ZLIB Deflator algorithm [36]. Table 3 shows the generated Huffman codes. Table 4 shows the size of the compressed difference codes. It also shows that ZLIB Deflator algorithm achieved better compression ratio than Huffman algorithm in human data, while Huffman algorithm achieved slightly better compression ratio in virus and mouse data.

Table 3. Huffman codes for difference operations.

| Relevant number | Human data | | Virus data | | Mouse data | |
|---|---|---|---|---|---|---|
| | Probability | Encoding value | Probability | Encoding value | Probability | Encoding value |
| 1 | 0.0011 | 11101 | 0.0641 | 110 | 0.0718 | 0011 |
| 2 | 0.0942 | 10 | 0.0653 | 101 | 0.0906 | 0010 |
| 3 | 0.0021 | 1111 | 0.0550 | 111 | 0.0483 | 0111 |
| 4 | 0.8957 | 0 | 0.7218 | 0 | 0.0696 | 0110 |
| 5 | 0.0002 | 1110001 | 0.0237 | 10001 | 0.191 | 10 |
| 6 | 0.0007 | 111001 | 0.025 | 10000 | 0.1849 | 11 |
| 7 | 0.005 | 110 | 0.0221 | 10011 | 0.1595 | 010 |
| 8 | 0.0005 | 1110000 | 0.0225 | 10010 | 0.1839 | 000 |

Table 4. Size of compressed differences for different compression algorithms

| Dataset | Uncompressed size | Huffman algorithm | ZLIB Deflator algorithm |
|---|---|---|---|
| Human | 1.08MB | 156.6KB | 35.6KB |
| Virus | 374.7KB | 81.6KB | 84.3KB |
| Mouse | 11.8KB | 4.3KB | 5.2KB |

Difference locations are compressed using ZLIB Deflator algorithm. Table 5 summarizes the compression ratio and space saving of our algorithm on the three datasets. The results confirm that mitochondria dataset that occupies 56MB size in GenBank can be stored using only 294.3KB, corresponding to 195-fold compression rate. While virus dataset that occupies 601KB size in GenBank can be stored using only 212.9KB, corresponding to 3-fold level of compression. Mouse datasets that occupies 106KB size in GenBank can be stored using only 9.6KB, corresponding to 11-fold level of compression. It should be noted that compression of mitochondria dataset is better than virus and mouse database because it has high similarity between genomic sequences.

Table 5. Comparison of compression ratios and space saving

| Dataset | Uncompressed size | Differences | Locations | Compression ratio | Space saving |
|---|---|---|---|---|---|
| Human | 56MB | 35.6KB | 258.7KB | 0.005 | 99.4% |
| Virus | 601KB | 81.6KB | 131.3KB | 0.354 | 64.6% |
| Mouse | 106KB | 4.3KB | 5.3KB | 0.09 | 91.0% |

The most popular algorithms, including specific compression algorithms: DNACompress [12], XM [13], and difference compression scheme: DNAEncodeWG [27], are not suitable for compression of entire databases. We believe that DNAEncodeWG is superior for a single sequence. It should be mentioned that Brandon et al. algorithm [25], was done by using extra reference sequence and based on statistics of the sequence set that are not valid for other sets. The compression result of our algorithm is better than C. Wang [26] that achieved 159-fold compression on Korean personal genome. Our results confirm that the proposed method compresses a 56MB human data file to 294.4KB, with a space savings of 99.4%. For virus data, a 601KB file was compressed to 212.9KB, with a space savings of 64.6%. For mouse data, a 106KB file was compressed to 9.6KB, with a space savings of 91%.

As can be observed, high similarity between genomic sequences is essential to improve the compression. For each dataset, the compression ratio achieved is a measure of the similarity between its sequences. In other words, it is possible to measure how much information one sequence gives about the other.

An important point should be mentioned here is that the proposed method does not require a priori knowledge about that statistics of the sequence set. This will be a big advantage when the updated sequences have different statistics.

The execution time of the differential compression is depending on the length of the difference sequence needed to be compressed. As the difference sequence length increases, the execution time increases. On the other hand, decoding speed is similar since both encoder and decoder do essentially the same computation. Therefore, our current implementation has been optimized for encoding and decoding speed, when the sequence set is under heavy usage.

The differential compression method has been used for lossless compression. Through our investigation, we have found that the use of this method could open new frontiers in quickly identifying unknown sequence related to the set of sequences. This method can potentially be modified to include variable references instead of using a single reference to improve the difference compression. It also extends to update the set with new similar sequences. Further investigation of the new method is needed to further assess its practical value. Moreover, when individuals have complete genome sequences available as part of their personal health records, the focus will shift to difference sequence-level compression. It seems likely that improvement of difference compression will continue to be important to advance knowledge of human genetic variation, and is the pressing problem faced by researchers today.

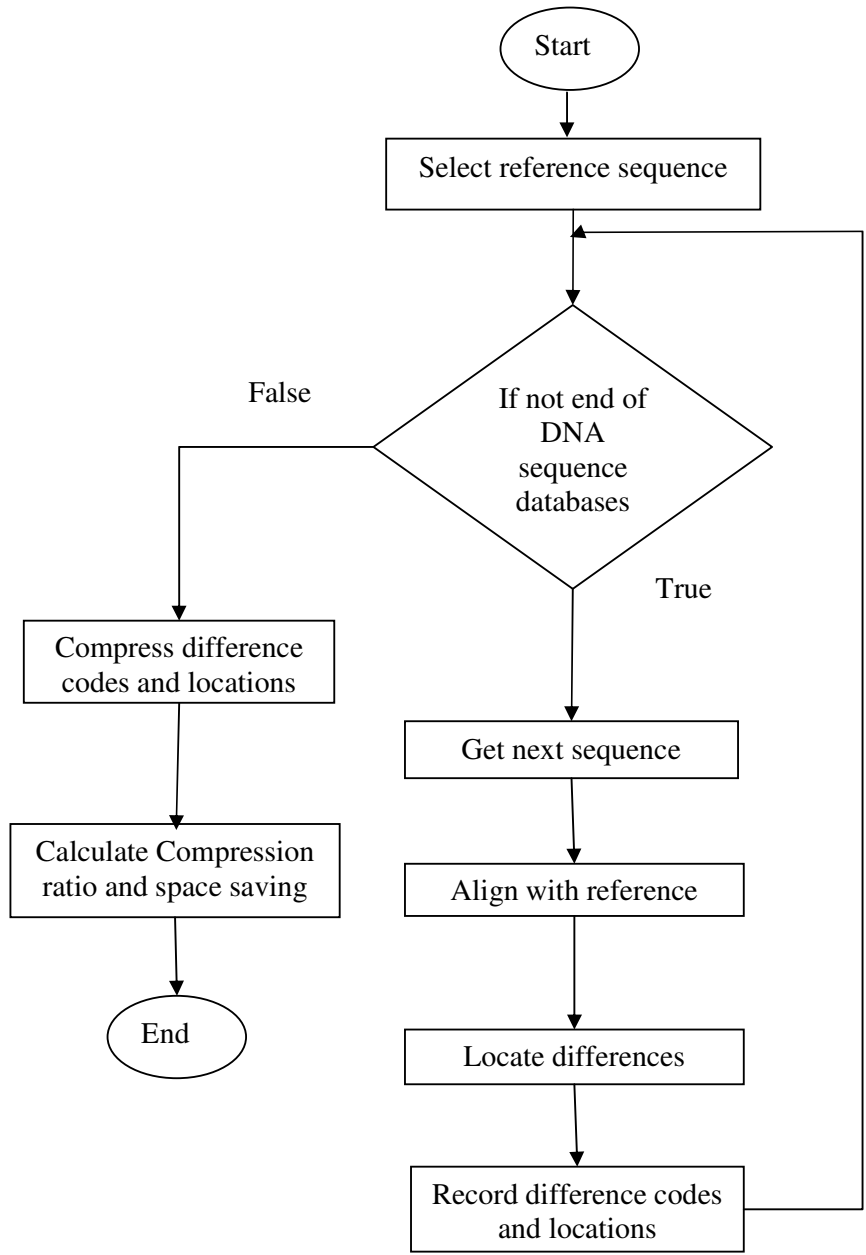

Figure 1. Architecture of the differential compression algorithm

# 4. CONCLUSIONS

The differential compression method that is based on the differences and locations of differences between each sequence in a genomic dataset and its reference sequence is presented. This method is simple, universal which does not depend on the statistics of the dataset. and could achieve up to 195-fold compression. Compression methods are undergoing rapid development making it tempting to store sequencing data for long periods of time so that the data can be re-analyzed with the latest techniques. The challenging open research problems, huge influx of data, and rapidly improving analysis techniques have created the need to store and transfer very large volumes of data. More work is needed to select the prefect reference for huge data to improve difference compression generation, and to further investigate the use of the new method in practical genomic applications.